\documentclass[onecolumn,12pt]{asme2ej}

\usepackage{graphicx}
\usepackage{amsxtra}
\usepackage{amsthm}
\usepackage{amssymb}
\usepackage{bm}
\usepackage{amsfonts}

\title{Discussion: ``Bayesian Optimal Design of Experiments for Inferring the Statistical Expectation of Expensive Black-Box Functions" (Pandita, P., Bilionis, I., and Panchal, J., 2019. ASME. J. Mech. Des. 141(10): 101404)}

\author{Xianliang Gong
    \affiliation{
	Department of Naval Architecture and Marine Engineering\\
	University of Michigan\\
	Ann Arbor, MI 48109\\
    Email: xlgong@umich.edu
    }	
}

\author{Yulin Pan
    \affiliation{
	Department of Naval Architecture and Marine Engineering\\
	University of Michigan\\
	Ann Arbor, MI 48109\\
    Email: yulinpan@umich.edu
    }	
}

\begin{document}

\maketitle    

\section{Introduction}  
In \cite{pandita2019bayesian}, the authors developed a sequential Bayesian optimal design framework to estimate the statistical expectation of a black-box function $f(\mathbf{x}): \mathbb{R}^{d} \to \mathbb{R}$. Let $\mathbf{x} \sim p(\mathbf{x})$ with $p(\mathbf{x})$ the probability distribution of the input $\mathbf{x}$, the statistical expectation is then defined as:
\begin{equation}
    q = \int f(\mathbf{x})p(\mathbf{x}){\rm{d}}\mathbf{x}.
\label{expectation}
\end{equation}
The function  $f(\mathbf{x})$ is not known a priori but can be evaluated at arbitrary $\mathbf{x}$ with Gaussian noise of variance $\sigma^2$:
\begin{equation}
    y = f(\mathbf{x}) + \epsilon, \quad \quad \quad \epsilon \sim \mathcal{N}(0, \sigma^2).
\label{noise}
\end{equation}

Based on the Gaussian process surrogate learned from the available samples $\mathbf{D}_n = \{\mathbf{X}_n, \mathbf{Y}_n\}$, i.e., $f(\mathbf{x}) | \mathbf{D}_n  \sim \mathcal{GP}(m_n(\mathbf{x}),k_n(\mathbf{x}, \mathbf{x}'))$, the next-best sample is chosen by maximizing the information-based acquisition $G(\tilde{\mathbf{x}})$: 
\begin{equation}
    \mathbf{x}_{n+1} = {\rm{argmax}}_{\tilde{\mathbf{x}}} \; G(\tilde{\mathbf{x}}),
\label{opt}
\end{equation}
where $G(\tilde{\mathbf{x}})$ computes the information gain of adding a sample at  $\tilde{\mathbf{x}}$, i.e. the expected KL divergence between the current estimation $p(q|\mathbf{D}_n)$ and the hypothetical nest-step estimation $p(q|\mathbf{D}_n, \tilde{\mathbf{x}},\tilde{y})$ (where $\tilde{y}$ follows the distribution of $\mathcal{N}(m_n(\tilde{\mathbf{x}}),k_n(\tilde{\mathbf{x}}, \tilde{\mathbf{x}}) + \sigma^2)$): 
\begin{align}
    G(\tilde{\mathbf{x}}) & = \mathbb{E}_{\tilde{y}} \Big[{\rm{KL}}\big(p(q|\mathbf{D}_n, \tilde{\mathbf{x}},\tilde{y})|p(q|\mathbf{D}_n)\big) \Big] 
\nonumber \\ 
    & = \iint p(q|\mathbf{D}_n, \tilde{\mathbf{x}},\tilde{y}) \log \frac{p(q|\mathbf{D}_n, \tilde{\mathbf{x}},\tilde{y})}{p(q|\mathbf{D}_n)} {\rm{{\rm{d}}}}q \; p(\tilde{y}|\tilde{\mathbf{x}},\mathbf{D}_n) {\rm{{\rm{d}}}} \tilde{y}. 
\label{IG}
\end{align}
It is noted that $G(\tilde{\mathbf{x}})$ also depends on the hyperparameter $\mathbf{\theta}$ in the learned Gaussian process $f(\mathbf{x}) | \mathbf{D}_n$. We neglect this dependence for simplicity, which does not affect the main derivation.  

As a major contribution of the discussed paper, the authors simplified the information-based acquisition as Eq. (30) in \cite{pandita2019bayesian}:
\begin{equation}
    G(\tilde{\mathbf{x}}) = \log (\frac{\sigma_1}{\sigma_2(\tilde{\mathbf{x}})}) + \frac{1}{2}\frac{\sigma^2_2(\tilde{\mathbf{x}})}{\sigma^2_1} - \frac{1}{2} + \frac{1}{2}\frac{v(\tilde{\mathbf{x}})^2}{\sigma^2_1 (\sigma^2_n(\tilde{\mathbf{x}}) + \sigma^2)}, 
\label{original}
\end{equation}
where $\sigma_1^2$ and $\sigma_2^2$ are respectively the variances of current estimation and hypothetical nest-step estimation of $q$; $v(\tilde{\mathbf{x}}) = \int  k_n(\tilde{\mathbf{x}},\mathbf{x}) p(\mathbf{x}) {\rm{d}} \mathbf{x}$ and $\sigma^2_n(\tilde{\mathbf{x}}) = k_n(\tilde{\mathbf{x}},\tilde{\mathbf{x}})$. Furthermore, for numerical computation of \eqref{original}, the authors developed analytical formula for each involved quantity (important for high-dimensional computation) under uniform distribution of $\mathbf{x}$.

The purpose of our discussion is to show the following two critical points:
\begin{enumerate}
    \item The last three terms of \eqref{original} always add up to zero, leaving a concise form with a much more intuitive interpretation of the acquisition.
\newline
    \item The analytical computation of \eqref{original} can be generalized to arbitrary input distribution of $\mathbf{x}$, greatly broadening the application of the developed framework.
\end{enumerate}
\noindent These two points are discussed respectively in \S2 and \S3.

\section{Derivation of the simplified acquisition $G(\tilde{\mathbf{x}})$}
To simplify Eq. \eqref{IG}, we first notice that $q|\mathbf{D}_n$ follows a Gaussian distribution with mean $\mu_1$ and variance $\sigma_1^2$:
\begin{align}
    & \quad \quad \quad \quad \quad \quad p(q|\mathbf{D}_n) = \mathcal{N}(q;\mu_1, \sigma_1^2), 
\label{current} \\
    \mu_1 &= \mathbb{E}\big[\int {f}(\mathbf{x})p(\mathbf{x}){\rm{d}}\mathbf{x} |\mathbf{D}_n \big] 
\nonumber \\
     & = \int m_n(\mathbf{x})p(\mathbf{x}){\rm{{\rm{d}}}} \mathbf{x}, 
\label{current_mean} \\
     \sigma_1^2 & = \mathbb{E}\Big[\big(\int {f}(\mathbf{x})p(\mathbf{x}){\rm{{\rm{d}}}}\mathbf{x}\big)^2 |\mathbf{D}_n \Big] -  (\mathbb{E}\Big[\big(\int {f}_n(\mathbf{x})p(\mathbf{x}){\rm{{\rm{d}}}}\mathbf{x}\big) |\mathbf{D}_n \Big])^2
\nonumber \\
     & = \iint k_n(\mathbf{x}, \mathbf{x}') p(\mathbf{x}) p(\mathbf{x}') {\rm{{\rm{d}}}} \mathbf{x}' {\rm{{\rm{d}}}}\mathbf{x}. 
\label{current_variance}
\end{align}
After adding one hypothetical sample $\{\tilde{\mathbf{x}},\tilde{y}\}$, the function follows an updated surrogate ${f}(x)|\mathbf{D}_n, \tilde{\mathbf{x}},\tilde{y}  \sim \mathcal{GP}(m_{n+1}(\mathbf{x}),k_{n+1}(\mathbf{x}, \mathbf{x}'))$ with
\begin{align}
    m_{n+1}(\mathbf{x}) & = m_{n}(\mathbf{x}) +  \frac{k_n(\tilde{\mathbf{x}},\mathbf{x})(\tilde{y}-m_n(\tilde{\mathbf{x}}))}{k_n(\tilde{\mathbf{x}}, \tilde{\mathbf{x}}) + \sigma^2},
\\
    k_{n+1}(\mathbf{x}, \mathbf{x}') & = k_n(\mathbf{x}, \mathbf{x}') -
    \frac{ k_n(\tilde{\mathbf{x}},\mathbf{x})k_n(\mathbf{x'}, \tilde{\mathbf{x}})}{k_n(\tilde{\mathbf{x}}, \tilde{\mathbf{x}}) + \sigma^2} .
\end{align}
The quantity $q| \mathbf{D}_n,  \tilde{\mathbf{x}},\tilde{y}$ can then be represented by another Gaussian with mean $\mu_2$ and variance $\sigma_2^2$:
\begin{align}
    & \quad \quad \quad \quad \quad \quad p(q| \mathbf{D}_n,  \tilde{\mathbf{x}},\tilde{y}) = \mathcal{N}(q;\mu_2 (\tilde{\mathbf{x}},\tilde{y}), \sigma_2^2 (\tilde{\mathbf{x}})) 
\label{next} ,\\
    \mu_2 (\tilde{\mathbf{x}},\tilde{y}) & = \mathbb{E} \big[\int {f}(\mathbf{x})p(\mathbf{x}){\rm{d}}\mathbf{x} |\mathbf{D}_n, \tilde{\mathbf{x}},\tilde{y} \big] 
\nonumber\\
    & = \int m_{n+1}(\mathbf{x})p(\mathbf{x}){\rm{{\rm{d}}}} \mathbf{x}, 
\nonumber\\
     & = \mu_1 + \frac{ \int k_n(\tilde{\mathbf{x}},\mathbf{x}) p(\mathbf{x}) {\rm{d}} \mathbf{x} }{k_n(\tilde{\mathbf{x}}, \tilde{\mathbf{x}}) + \sigma^2} (\tilde{y}-m_n(\tilde{\mathbf{x}})), 
\label{next_mean}\\
     \sigma_2^2 (\tilde{\mathbf{x}}) & = \mathbb{E}\Big[\big(\int {f}(\mathbf{x})p(\mathbf{x}){\rm{{\rm{d}}}}\mathbf{x}\big)^2 |\mathbf{D}_n, \tilde{\mathbf{x}},\tilde{y} \Big] -  (\mathbb{E}\Big[\big(\int {f}_n(\mathbf{x})p(\mathbf{x}){\rm{{\rm{d}}}}\mathbf{x}\big) |\mathbf{D}_n, \tilde{\mathbf{x}},\tilde{y} \Big])^2 
\nonumber \\
    & = \iint k_{n+1}(\mathbf{x}, \mathbf{x}') p(\mathbf{x}) p(\mathbf{x}') {\rm{{\rm{d}}}} \mathbf{x}' {\rm{{\rm{d}}}}\mathbf{x}
\nonumber \\
     & = \sigma_1^2 - \frac{ (\int k_n(\tilde{\mathbf{x}},\mathbf{x}) p(\mathbf{x}){\rm{d}} \mathbf{x})^2}{k_n(\tilde{\mathbf{x}}, \tilde{\mathbf{x}}) + \sigma^2} 
\label{next_variance},
\end{align}

We note that Eq. \eqref{current_mean}, Eq. \eqref{current_variance}, Eq. \eqref{next_mean}, and Eq. \eqref{next_variance} are respectively intermediate steps of Eq. (19), Eq. (21), Eq. (26) and Eq. (28) in the discussed paper. Substitute Eq. \eqref{current} and Eq. \eqref{next} into Eq. \eqref{IG}, one can obtain:
\begin{align}
       G(\tilde{\mathbf{x}})  & =  \iint p(q|\mathbf{D}_n, \tilde{\mathbf{x}},\tilde{y}) \log \frac{p(q|\mathbf{D}_n, \tilde{\mathbf{x}},\tilde{y})}{p(q|\mathbf{D}_n)} dq \; p(\tilde{y}|\tilde{\mathbf{x}},\mathbf{D}_n)) {\rm{{\rm{d}}}} \tilde{y} 
\nonumber \\ 
     & = \int (\log (\frac{\sigma_1}{\sigma_2(\tilde{\mathbf{x}})}) + \frac{\sigma^2_2(\tilde{\mathbf{x}})}{2 \sigma^2_1} + \frac{(\mu_2(\tilde{\mathbf{x}},\tilde{y})-\mu_1)^2}{2 \sigma^2_1} - \frac{1}{2}) p(\tilde{y}|\tilde{\mathbf{x}},\mathbf{D}_n) {\rm{{\rm{d}}}} \tilde{y} 
\nonumber \\
      & = \log (\frac{\sigma_1}{\sigma_2(\tilde{\mathbf{x}})}) +\frac{1}{2 \sigma^2_1} ( \int (\mu_2(\tilde{\mathbf{x}},\tilde{y})-\mu_1)^2  p(\tilde{y}|\tilde{\mathbf{x}},\mathbf{D}_n)  {\rm{{\rm{d}}}} \tilde{y}  + \sigma^2_2(\tilde{\mathbf{x}}) - \sigma^2_1 ) 
\nonumber \\
      & = \log (\frac{\sigma_1}{\sigma_2(\tilde{\mathbf{x}})}) + \frac{1}{2 \sigma^2_1}(\frac{ (\int  k_n(\tilde{\mathbf{x}},\mathbf{x}) p(\mathbf{x}) {\rm{d}} \mathbf{x})^2 }{k_n(\tilde{\mathbf{x}}, \tilde{\mathbf{x}}) + \sigma^2}  + \sigma^2_2(\tilde{\mathbf{x}}) - \sigma^2_1 )  
\label{original2}\\
      & = \log (\frac{\sigma_1}{\sigma_2(\tilde{\mathbf{x}})}).
\label{opt_S}
\end{align}
where Eq. \eqref{original2} is exactly Eq. \eqref{original} (or Eq. (30) in discussed paper). The fact that the last three terms of Eq. \eqref{original2} sum up to zero is a direct result of \eqref{next_variance}.

The advantage of having a simplified form \eqref{opt_S} is that the optimization \eqref{opt} yields a much more intuitive physical interpretation. Since $\sigma_1$ does not depend on $\tilde{\mathbf{x}}$, \eqref{opt} can be reformulated as
\begin{equation}
     \mathbf{x}_{n+1} = {\rm{argmin}}_{\tilde{\mathbf{x}}} \; \sigma^2_2(\tilde{\mathbf{x}}),
\label{acq_sigma}
\end{equation}
which selects the next-best sample minimizing the expected variance of $q$. Similar optimization criterion is also used in \cite{hu2016global} and \cite{blanchard2020output} for the purpose of computing the extreme-event probability.

Another alternative interpretation can be obtained by writing \eqref{opt} as
\begin{align}
    \mathbf{x}_{n+1} 
    & \equiv {\rm{argmax}}_{\tilde{\mathbf{x}}} \; \sigma_1^2 - \sigma^2_2(\tilde{\mathbf{x}})
\nonumber \\ 
    &  \equiv {\rm{argmax}}_{\tilde{\mathbf{x}}} \; \big(\frac{ \int  k_n(\tilde{\mathbf{x}},\mathbf{x}) p(\mathbf{x}){\rm{d}} \mathbf{x}}{(k_n(\tilde{\mathbf{x}}, \tilde{\mathbf{x}}) + \sigma^2)^{\frac{1}{2}}} \big)^2 
\label{acq0}\\
    & \equiv {\rm{argmax}}_{\tilde{\mathbf{x}}} \big( \int  \rho_{y|\mathbf{D}_n}(\tilde{\mathbf{x}},\mathbf{x}) (k_n(\mathbf{x},\mathbf{x})+ \sigma^2)^{\frac{1}{2}}  p(\mathbf{x}){\rm{d}} \mathbf{x} \big)^2 ,
\label{acq1} 
\end{align}
where \eqref{acq0} is a result of \eqref{next_variance}, and  $\rho_{y|\mathbf{D}_n}(\tilde{\mathbf{x}},\mathbf{x}) =  k_n(\tilde{\mathbf{x}},\mathbf{x})/\big((k_n(\tilde{\mathbf{x}}, \tilde{\mathbf{x}}) + \sigma^2)(k_n(\mathbf{x},\mathbf{x})+ \sigma^2)\big)^{\frac{1}{2}}$ is the correlation of $y$ for two inputs $\tilde{\mathbf{x}}$ and $\mathbf{x}$. Eq. \eqref{acq1} can be interpreted as to select the next sample which has overall most (weighted) correlation with all $\mathbf{x}$. 

We finally remark that the above derivation is for given hyperparamter values $\theta$ in $f(\mathbf{x}) | \mathbf{D}_n$. This is consistent with the Bayesian approach where the optimal values of $\theta$ are chosen from maximizing the likelihood function. However, the discussed paper used a different approach by sampling a distribution of $\theta$ and computed $G(\tilde{\mathbf{x}})$ as an average of the sampling. In the latter case, the above analysis should be likewise considered in a slightly different way, i.e., Eq. \eqref{acq_sigma} should be considered as maximization of the multiplication of $\sigma_2^2$ from all samples of $\{\mathbf{\theta}^{(i)}\}_{i=1}^{s}$:
\begin{align}
    \mathbf{x}_{n+1} & = {\rm{argmax}}_{\tilde{\mathbf{x}}} \; \frac{1}{s} \sum_{i=1}^s G(\tilde{\mathbf{x}},\mathbf{\theta}^{(i)})
\nonumber \\
    & \equiv {\rm{argmax}}_{\tilde{\mathbf{x}}} \; \frac{1}{s} \sum_{i=1}^s \log (\frac{\sigma_1(\mathbf{\theta}^{(i)})}{\sigma_2(\tilde{\mathbf{x}},\mathbf{\theta}^{(i)})}) 
\nonumber \\
    & \equiv {\rm{argmin}}_{\tilde{\mathbf{x}}} \; \frac{1}{s} \sum_{i=1}^s \log \sigma_2(\tilde{\mathbf{x}},\mathbf{\theta}^{(i)})
\nonumber \\
    &  \equiv {\rm{argmin}}_{\tilde{\mathbf{x}}} \; \prod_{i=1}^{s} \sigma^2_2(\tilde{\mathbf{x}}, \mathbf{\theta}^{(i)}).
\label{acq2} 
\end{align}

\section{Analytical computation of $G(\mathbf{x})$ for arbitrary input distribution $p(\mathbf{x})$}
In the computation of $G(\mathbf{x})$ in the form of Eq. \eqref{acq0}, the most heavy computation involved is the integral $\int  k_n(\tilde{\mathbf{x}},\mathbf{x}) p(\mathbf{x}){\rm{d}} \mathbf{x}$ (which is prohibitive in high-dimensional problem if direct integration is performed). Following the discussed paper, the integral can be reformulated as 
\begin{equation}
    \int  k_n(\tilde{\mathbf{x}},\mathbf{x}) p(\mathbf{x}){\rm{d}} \mathbf{x} = 
    \mathcal{K}(\tilde{\mathbf{x}}) - \mathbf{k}(\tilde{\mathbf{x}},\mathbf{X}_n)\big(\mathbf{K}(\mathbf{X}_n, \mathbf{X}_n) + \sigma^2 \mathbf{I}_n\big)^{-1}\mathcal{K}(\mathbf{X}_n) 
\end{equation}

where 
\begin{align}
    \mathcal{K}(\mathbf{x}) & = \int k(\mathbf{x},\mathbf{x}')p(\mathbf{x}') {\rm{d}} \mathbf{x}'
\label{kappa}\\
     k(\mathbf{x},\mathbf{x}') & = s^2 \exp{\big(-\frac{1}{2}(\mathbf{x}- \mathbf{x}')^T \Lambda^{-1}(\mathbf{x}- \mathbf{x}')\big)},
\label{RBF}
\end{align}
with $s$ and $\Lambda$ involving hyperparameters of the kernel function (with either optimized values from training or selected values as in \cite{pandita2019bayesian}). 

The main computation is then \eqref{kappa}, for which the authors of the discussed paper addressed the situation of uniform $p(\mathbf{x})$. To generalize the formulation to arbitrary $p(\mathbf{x})$, we can approximate $p(\mathbf{x}) $ with the Gaussian mixture model (as a universal approximator of distributions \cite{Goodfellow-et-al-2016}) :
\begin{equation}
    p(\mathbf{x}) \approx \sum_{i=1}^{n_{GMM}} \alpha_i \mathcal{N}(\mathbf{x};\mathbf{w}_i,\Sigma_i).
\label{gmm}
\end{equation}
Eq. \eqref{kappa} can then be formulated as:
\begin{align}
    \mathcal{K}(\mathbf{x}) & \approx \sum_{i=1}^{n_{GMM}} \alpha_i \int k(\mathbf{x},\mathbf{x}') \mathcal{N}(\mathbf{x}';\mathbf{w}_i,\Sigma_i) {\rm{d}} \mathbf{x}' 
\nonumber \\
    & = \sum_{i=1}^{n_{GMM}} \alpha_i|\Sigma_i \Lambda^{-1} + {\rm{I}}|^{-\frac{1}{2}}k(\mathbf{x},\mathbf{w}_i;\Sigma_i + \Lambda),
\label{kappa_gmm}
\end{align}
which yields an analytical computation. In practice, the number of mixtures $n_{GMM}$ is determined by the complexity of the input distributions, but any distribution of $p(\mathbf{x})$ can be approximated in such a way.

Finally, in computing $G(\tilde{\mathbf{x}})$ in the form of \eqref{acq2}, computation of $\sigma^2_1$ in \eqref{current_variance} is necessary. This can also be generalized for arbitrary $p(\mathbf{x})$ using the Gaussian mixture model as follows:
\begin{align}
    \sigma^2_1 = & \iint k_n(\mathbf{x}, \mathbf{x}') p(\mathbf{x}) p(\mathbf{x}') {\rm{{\rm{d}}}} \mathbf{x}' {\rm{{\rm{d}}}}\mathbf{x} 
\nonumber \\
     = & \iint k(\mathbf{x}, \mathbf{x}') p(\mathbf{x}) p(\mathbf{x}') {\rm{{\rm{d}}}} \mathbf{x}' {\rm{{\rm{d}}}}\mathbf{x} 
     - \int \mathbf{k}(\mathbf{x},\mathbf{X}_n) p(\mathbf{x}) {\rm{d}} \mathbf{x} \;
     \big(\mathbf{K}(\mathbf{X}_n, \mathbf{X}_n) + \sigma^2 \mathbf{I}_n \big)^{-1} \int \mathbf{k}(\mathbf{X}_n, \mathbf{x}')
     p(\mathbf{x}') {\rm{d}} \mathbf{x}' 
\nonumber \\
   = & \sum_{i=1}^{n_{GMM}} \sum_{j=1}^{n_{GMM}}  \alpha_i  \alpha_j
   |\Lambda|^{1/2} |\Lambda + \Sigma_i + \Sigma_j|^{-1/2} k(\mathbf{w}_i, \mathbf{w}_j;\Lambda + \Sigma_i + \Sigma_j)
\nonumber \\
    & \quad \quad \quad \quad \quad \quad \quad \quad \quad \quad
    \quad \quad \quad \quad \quad
    -\mathcal{K}(\mathbf{X}_n)^T \big(\mathbf{K}(\mathbf{X}_n, \mathbf{X}_n) + \sigma^2 \mathbf{I}_n \big)^{-1}\mathcal{K}(\mathbf{X}_n).
\end{align}



\bibliographystyle{asmems4}
\bibliography{asme2e}

\end{document}